\documentclass[12pt]{article}

\usepackage{amsmath,amssymb}
\usepackage{graphicx}
\usepackage{caption}
\usepackage{color}
\usepackage{dcolumn}
\usepackage{bm}
\usepackage{enumerate}
\usepackage[numbers,super,comma,sort&compress]{natbib}
\usepackage{tabularx}
\usepackage{xfrac}

\definecolor{background-color}{gray}{0.98}

\newcommand{\refeq}[1]{equation~(\ref{#1})}
\newcommand{\reffig}[1]{Figure~\ref{#1}}

\newcommand{\mysection}[1]{\section*{\sffamily \Large #1}}
\newcommand{\mysubsection}[1]{\subsection*{\sffamily \large #1}}
\newcommand{\mysubsubsection}[1]{\subsubsection*{\sffamily \normalsize #1}}
\newcommand{\ie}{\emph{i.e.}}
\newcommand{\dd}{\mathrm{d}}

\begin{document}

\title{Gaussian Approximation Potentials: a brief tutorial introduction}
\author{Albert P. Bart\'ok and G\'abor Cs\'anyi\\
\em Engineering Laboratory, University of Cambridge, Trumpington Street,\\
\em Cambridge, CB2 1PZ, United Kingdom}

\maketitle

\begin{abstract}
 We present a swift walk-through of our recent work that uses machine learning to fit interatomic potentials based on quantum mechanical data.  We describe our Gaussian Approximation Potentials (GAP) framework, discuss a variety of descriptors,  how to train the model on total energies and derivatives and the simultaneous use of multiple models of different complexity. We also show a small example using     QUIP, the software sandbox implementation of GAP that is available for non-commercial use.
\end{abstract}

\clearpage

  \makeatletter
  \renewcommand\@biblabel[1]{#1.}
  \makeatother

\bibliographystyle{apsrev}

\renewcommand{\baselinestretch}{1.5}
\normalsize

\clearpage

\mysection{INTRODUCTION} 
Molecular scale simulation is a mature field with a wide range of electronic
structure methods that approximate the solution of the Schr\"odinger equation in a systematic fashion. For larger scale computations empirical interatomic potentials  are used, which are nowadays fit to data generated by electronic structure models. Together these play a significant r\^ole in understanding processes on the microscopic level, complementing experiment and theory. Computer simulations are regularly  used to interpret experimental results and to predict properties of materials.

The power of atomistic simulations would be enormously enhanced if the interatomic potentials used to simulate materials were not limited by their simple empirical functional forms but accurately approached the Born-Oppenheimer potential energy surface, similarly to the case of small molecules for which quantum chemists have been fitting accurate potential energy surfaces for decades. The challenge in the materials field is that rather than fitting the total energy of a fixed number of atoms, the task is to find a unique local  functional that describes the energy of a single atom or bond given its neighbour environment. This local energy function must naturally allow for bond forming and bond breaking, i.e. the change in the number and identity of the atoms comprising the neighbour environment.

A number of  groups---many of them  contributing to the present volume---have
started research programmes to address this problem using  advances in the
synthetic understanding that recently emerged in statistics and   machine
learning.\cite{Behler:2007th,Behler:2011kh,Rupp:2012kx,Brown:2003tu,Hansen:2013dp,Handley:2010ft} These fast-growing fields are concerned with classification,
regression and probability density estimation on large and noisy data sets, and
also with finding suitable variable transformations that allow increased
performance in these tasks. There are a number of closely related computational
frameworks that are widely used, including artificial neural networks,
stochastic processes (e.g. Gaussian processes) and regularised non-parametric
optimisation. In this tutorial introduction we focus on a particular exposition
that allows  a succinct presentation of the formalism and how it can be brought
to bear on the problem of fitting potential energy surfaces for materials based
on data computed by electronic structure methods. For detailed derivations of
the necessary fundamental results we refer the reader to the machine learning
and statistics literature\cite{Mackay:tw,Scholkopf:2001}.

\begin{table}[h]
\begin{tabularx}{\linewidth}{|l|X|}\hline
\em neighbourhood & Set of nearby atoms whose positions constitute the input to the local energy function evaluated for a given atom.\\ \hline
\em descriptors & Transformation of the positions of atoms in the neighbourhood, obeying the desired symmetries of the energy function. Also called {\em features}.\\ \hline
\em kernel & Similarity measure between two neighbourhoods, equivalent to the {\em covariance} of the corresponding two local energy values.\\ \hline
\end{tabularx}
\caption{\label{tbl:dict}Definition of central concepts used in fitting accurate potentials for materials. }
\end{table}

\mysection{METHODOLOGY}
The hallmark of an interatomic potential is that the total energy, $E$,  of a set of atoms is written  as a sum of range-separated terms,
\begin{equation}\label{eq:totale_full_mbe}
E=\sum_\alpha \sum_{i \in \alpha} \varepsilon^\alpha_i + \textrm{long range contributions}
\end{equation}
where $\varepsilon^\alpha_i$ are   local energy functionals with compact support within a radius $r_\textrm{cut}$, and by ``long range contributions'' we mean
electrostatics including polarisability, van der Waals interactions etc. This is an uncontrolled approximation, since there is nothing about the Schr\"odinger equation that tells us {\em a priori} that its solutions can be written in this form: the level of accuracy and its applicability in any particular situation has to be tested by numerical experiments. 
The index $\alpha$ denotes the type of contribution: the arguments of a local energy term  may be any suitable {\em descriptors},  e.g. atom-pair distances, bond angles, or indeed the complete atomic environment, and the index $i$ counts the instances of these terms in a particular configuration, e.g.  all bonds for a pair term, all angles for a three-body angle-dependent term, or all
atoms for an atom-centered term. We can think of descriptors as functions that transform the Cartesian coordinates of the  atoms
in the neighbourhood of a given atom.   

In this paper we will only discuss the local energy contribution, although it is clear that for many materials in which atoms acquire significant partial charges or have easily polarisable electrons it must be complemented by electrostatic and dispersion interactions. These long range terms can either remain completely empirical, but may also include parameters that are fitted to data using approaches similar to what are used for the local term.

\mysubsection{Gaussian Process Regression}

We first consider the case of a single type of local energy functional. Using a set of arbitrary basis functions $\{\phi_h\}_{h=1}^H$ that take as their arguments any descriptor  $\mathbf{d}_i$ of the neighbour environment of atom $i$,  we write the atomic energy $\varepsilon_i$ as
\begin{equation}
\varepsilon_i = \varepsilon(\mathbf{d}_i,\mathbf{w}) = \sum_h w_h \phi_h(\mathbf{d}_i)
\textrm{,}
\end{equation}
where $\mathbf{w}$ is a vector of   weights $w_h$ corresponding to the basis functions, to be determined by the fit.  If  the prior probability distribution of the weights is chosen to be Gaussian with zero mean, i.e. $P(\mathbf{w})=\mathrm{Normal}(\mathbf{w}; \mathbf{0}, \sigma_w \mathbf{I})$, 
the covariance of two atomic energies is
\begin{equation}\label{eq:atomic_variance}
\langle \varepsilon_i \varepsilon_j \rangle = \left \langle \sum_{h h'} w_h w_{h'} \phi_h(\mathbf{d}_i) \phi_{h'}(\mathbf{d}_j) \right \rangle = 
\sum_{h h'} \langle w_h w_{h'} \rangle \phi_h(\mathbf{d}_i) \phi_{h'}(\mathbf{d}_j) = \sigma_w^2 \sum_h \phi_h(\mathbf{d}_i) \phi_{h}(\mathbf{d}_j) 
\end{equation}
where we exploited that $\langle w_h w_{h'} \rangle = \delta_{hh'} \sigma_w^2$. 
The inner product of the basis functions in the last expression defines the \emph{kernel} or \emph{covariance function}
\begin{equation}
C(\mathbf{d}_i,\mathbf{d}_j) \equiv \sum_h \phi_h(\mathbf{d}_i) \phi_{h}(\mathbf{d}_j)
\textrm{.}
\end{equation}
  Kernel functions in this application are to be understood as similarity measures between two atomic neighbour environments. Every basis set induces a corresponding kernel, and as seen below, only the kernel is required for regression, we never need to construct a basis set in the space of descriptors explicitly.  General requirements on  kernel functions  are in the literature\cite{Mackay:tw,Williams:2007vz}.

Our goal is to predict the  energy of an arbitrary atomic configuration, based upon a data set of previous calculations.
For any set of microscopic observations $\mathbf{t}$---which could be the local atomic energies or the total energies of all atoms in a set of configurations---the covariance matrix is defined as $\mathbf{C} \equiv \langle \mathbf{t} \mathbf{t}^\top \rangle$, and its elements can be computed using the previously defined covariance function. The prior probability of observing  $\mathbf{t}$ is also Gaussian,
\begin{equation}
P(\mathbf{t}) = \mathrm{Normal}(\mathbf{t}; \mathbf{0}, \mathbf{C}) \propto \exp \left( -\frac{1}{2} \mathbf{t}^\top \mathbf{C}^{-1} \mathbf{t} \right)
\textrm{.}
\end{equation}
The predicted value  $y$ of a new test configuration, given previous observations  $\mathbf{t}$, has the probability distribution
\begin{equation}
P(y \, | \, \mathbf{t}) = \frac{P(\mathbf{t},y)}{P(\mathbf{t})}
\end{equation}
which is also Gaussian. We take  the mean of this distribution as the prediction, which can be expressed\cite{Mackay:tw} as
\begin{equation}
\label{eq:predict}
\bar{y} = \mathbf{k}^\top \mathbf{C}^{-1} \mathbf{t}
\end{equation}
where $\mathbf{k}$ is the covariance vector of function values: $\mathbf{k} \equiv \langle y \, \mathbf{t} \rangle$. This shows  the real power of the Gaussian process approach: the original basis functions we started with and their corresponding unknown weights  are never
required explicitly, the predictions only depend on the kernel function $C$ and the previous observations $\mathbf{t}$.

It can be shown\cite{Neal:1995ws} that a two-layer neural network with infinite number of hidden nodes and hyperbolic tangent switching function is equivalent to a Gaussian process with 
\begin{equation}
C(\mathbf{d}_i,\mathbf{d}_j) \propto V - |\mathbf{d}_i-\mathbf{d}_j|^2
\textrm{.}
\end{equation}
Extra layers in neural networks with more than two layers can be regarded as performing a nonlinear transformation on the input coordinates, before the output layers carry out the regression task.

Yet another equivalent approach for fitting functions is kernel ridge regression, where  the unknown function is expanded as a linear combination of radial basis functions\footnote{We note that kernel ridge regression is not limited to radial basis functions, any positive definite kernel may be used.\cite{Williams:2007vz}},
\begin{equation}
f(\mathbf{d})=\sum_i \alpha_i C(\mathbf{d},\mathbf{d}_i)
\textrm{,}
\end{equation}
and the weights $\boldsymbol{\alpha}$ are optimised by minimising the cost function
\begin{equation}
L=\sum_i (t_i - f(\mathbf{d_i}))^2 + \lambda || \boldsymbol{\alpha} ||^2
\textrm{.}
\end{equation}
If we define the norm as
\begin{equation}
 || \boldsymbol{\alpha} ||^2 = \boldsymbol{\alpha}^\top \mathbf{C} \, \boldsymbol{\alpha}
\textrm{,}\end{equation}
the predictions of kernel regression are also equivalent to those of the Gaussian process.
The kernel here has the dual role of defining both the basis functions and the norm of the weights in the loss function. 
\mysubsubsection{Total energies}
Atomic energies are  unavailable in quantum mechanical calculations, which only provide the total energy and its derivatives. From these, we have to predict the local energies. It is straightforward to modify \refeq{eq:atomic_variance} to express the covariance of the total energies of two set of atoms, $N$ and $M,$
\begin{multline}\label{eq:TE_covariance}
\langle E_N E_M \rangle = \left \langle \sum_{i \in N} \varepsilon(\mathbf{d}_i) \sum_{j \in M} \varepsilon(\mathbf{d}_j)\right \rangle = \left \langle \sum_{i \in N} \sum_{j \in M} \sum_{h h'} w_h w_{h'} \phi_h(\mathbf{d}_i) \phi_{h'}(\mathbf{d}_j) \right \rangle = \\
\sum_{i \in N} \sum_{j \in M} \sum_{h h'} \langle w_h w_{h'} \rangle \phi_h(\mathbf{d}_i) \phi_{h'}(\mathbf{d}_j) = 
\sigma_w^2 \sum_{i \in N} \sum_{j \in M} \sum_h \phi_h(\mathbf{d}_i) \phi_{h}(\mathbf{d}_j) =
\sigma_w^2 \sum_{i \in N} \sum_{j \in M} C(\mathbf{d}_i,\mathbf{d}_j)
\end{multline}
\mysubsubsection{Derivatives}
The total quantum mechanical energy of a configuration depends on the relative positions of the atoms and, in case of condensed systems, also the lattice parameters. Denoting a general coordinate by $\xi$, the partial derivative of the total energy is related to the force as
\begin{equation}
f_{k\alpha} =  -\frac{\partial E}{\partial r_{k\alpha}} = -\frac{\partial E}{\partial \xi}  \textrm{ if } \xi \equiv r_{k\alpha}
\end{equation}
or to the viral stress as
\begin{equation}
v_{\alpha\beta} =  \frac{\partial E}{\partial h_{\alpha\beta}} = \frac{\partial E}{\partial \xi}  \textrm{ if }
\xi \equiv h_{\alpha\beta}
\end{equation}
where $r_{k\alpha}$ is the $\alpha$-th component of the Cartesian coordinates of atom $k$ and $h_{\alpha\beta}$ is an element of the deformation matrix $\mathbf{H}$ of the lattice vectors. Differentiating \refeq{eq:TE_covariance} with respect to an arbitrary coordinate $\xi_k$ of configuration $N$ results in
\begin{equation}\label{eq:dTE_covariance}
\left \langle \frac{\partial  E_N}{\partial \xi_k} E_M \right \rangle =
\frac{\partial \langle E_N E_M \rangle}{\partial \xi_k} = 
\sigma_w^2 \sum_{i \in N} \sum_{j \in M} \nabla_{\mathbf{d}_i} C(\mathbf{d}_i,\mathbf{d}_j) \cdot
\frac{\partial \mathbf{d}_i}{\partial \xi_k}
\textrm{.}
\end{equation}
If $\xi_k$ is the $x$, $y$, or $z$ component of the position of atom $k$, $\frac{\partial \mathbf{d}_i}{\partial \xi_k}$ becomes exactly zero if the pair distance $|\mathbf{r}_i-\mathbf{r}_k|$ is beyond the cutoff of the environment, so the first sum need not be done over all atoms in the configuration. Similarly, the covariance of two derivative quantities may be written as
\begin{equation}\label{eq:d2TE_covariance}
\left \langle \frac{\partial  E_N}{\partial \xi_k} \frac{\partial  E_M}{\partial \chi_l}  \right \rangle =
\frac{\partial^2 \langle E_N E_M \rangle}{\partial \xi_k \partial \chi_l} = 
\sigma_w^2 \sum_{i \in N} \sum_{j \in M} \frac{\partial \mathbf{d}_i^\top}{\partial \xi_k}  (\nabla_{\mathbf{d}_i} C(\mathbf{d}_i,\mathbf{d}_j) \nabla_{\mathbf{d}_j}^\top)
\frac{\partial \mathbf{d}_j}{\partial \chi_l}
\textrm{,}
\end{equation}
where the elements of the Jacobian are 
\begin{equation}
(\nabla_{\mathbf{d}_i} C(\mathbf{d}_i,\mathbf{d}_j)\nabla_{\mathbf{d}_j}^\top)_{\alpha\beta} = 
\frac{\partial^2 C(\mathbf{d}_i,\mathbf{d}_j)}{\partial d_{i\alpha}\partial d_{j\beta}}
\end{equation}

The local energy $\varepsilon$ is still predicted by using \refeq{eq:predict}, but the elements of $\mathbf{y}$ are  total energies or derivative quantities, and the elements of the covariance matrix $\mathbf{C}$ are therefore computed by equations~(\ref{eq:TE_covariance}), (\ref{eq:dTE_covariance}) or (\ref{eq:d2TE_covariance}). The elements of $\mathbf{k}$ are the covariance between the local energy that we wish to predict and the data that we have available,  $\langle \varepsilon \, E \rangle$ or $\langle \varepsilon \, \sfrac{\partial E}{ \partial \xi} \rangle$ as appropriate.

\mysubsubsection{Multiple models}
Interactions in some atomistic systems might be partitioned using a many-body type expansion -- indeed, many traditional interatomic potentials are based on a few low-order contributions, such as two- and three-body energies\cite{STILLINGER:1985vx}. We now describe how such models can be fitted using Gaussian process regression. For example, truncating the local part of \refeq{eq:totale_full_mbe} at   three-body contributions, the total energy is approximated as
\begin{equation}
E = \sum_{p \,\in\, \textrm{ pairs}} \varepsilon^{(2)}_p +\sum_{t \,\in\, \textrm{ triplets}} \varepsilon^{(3)}_t
\end{equation}
where $\varepsilon^{(2)}$ and $\varepsilon^{(3)}$ are general two- and three-body energy functions, respectively, and  pairs and triplets in this context may refer to atoms as well as entire molecules. Two independent Gaussian processes are used,
\begin{align}
\varepsilon^{(2)}(\cdot \cdot) &= \sum_h w^{(2)}_h \phi^{(2)}_h(\cdot \cdot) \textrm{ }\\
\varepsilon^{(3)}(\therefore) &= \sum_h w^{(3)}_h \phi^{(3)}_h(\therefore) \textrm{,}
\end{align}
where $\cdot\cdot$ and $\therefore$ denotes generic geometric descriptors of pairs and triplets (in case of molecules, the descriptors need to describe the whole dimer and trimer configuration). The prior distributions of the two weight vectors are independent Gaussians, so the covariance of the total energy of two configurations $N$ and $M$ may be written as
\begin{equation}
\langle E_N E_M \rangle = \sigma^2_{w^{(2)}} \sum_{p \in \textrm{pairs}_N}\sum_{q \in \textrm{pairs}_M} C^{(2)}(p,q) +
\sigma^2_{w^{(3)}} \sum_{t \in \textrm{triplets}_N}\sum_{u \in \textrm{triplets}_M} C^{(3)}(t,u) \textrm{,}
\end{equation}
where we applied the same kernel trick as above and exploited that $\langle w^{(2)}_h w^{(3)}_{h'}\rangle = 0$ for any $h$ and $h'$. As the two-body terms can, in principle, be included in the three-body terms, splitting them appropriately might require setting the the variances of the two terms carefully. For example, if 80\% of the total interaction  energy is expected to be due to pair interactions, this information  can be built into the prior by using the ratio 
$\sigma_{w^{(2)}}:\sigma_{w^{(3)}}=4:1$.

\mysubsubsection{Compact support}
The local energy terms need to have compact support to be computationally efficient, and this is typically achieved by using an explicit spatial cutoff function. In machine learning models for materials, the  cutoff may be built into descriptors\cite{Bartok:2013cs}, so only neighbours within a predefined radial distance of the central atom are considered. Alternatively, cutoffs may be implemented in the kernels. Consider the pair energy model
\begin{equation}
\varepsilon^{(2)}(\cdot \cdot) =  f_\textrm{cut}(\cdot\cdot) \sum_h w^{(2)}_h \phi^{(2)}_h(\cdot \cdot)
\end{equation}
where $f_\textrm{cut}$ is defined such that it goes smoothly to zero  as a function of the geometric attributes of the pair (e.g.  as the distance between them approaches a limit, in case of a pair of atoms). The resulting covariance function is
\begin{equation}
\langle \varepsilon^{(2)}(p) \varepsilon^{(2)}(q) \rangle = \sigma^2_{w^{(2)}} C^{(2)}(p,q) f_\textrm{cut}(p) f_\textrm{cut}(q)
\textrm{.}
\end{equation}
In our implementation we use
\begin{equation}
f_\textrm{cut}(r) = \left\{
\begin{array}{c l}
1 & \quad \text{for } r \le r_\textrm{cut} - d \\
\left[ \cos \left( \pi\frac{r-r_\textrm{cut}+d}{d} \right) + 1 \right]/2 & \quad \text{for } r_\textrm{cut}-d < r \le r_\textrm{cut} \\
0 & \quad \text{for } r > r_\textrm{cut}
\end{array}
\right.
\end{equation}
as the cutoff function, where $d$ is a parameter that determines the width of the cutoff region. There is some freedom in choosing a numerical value for $d$, but two criteria has to be considered: the covariance should change smoothly when two atoms become connected, and minimising the spurious effect of the cutoff transition on the derivatives. We typically use $d=1\,\textrm{\AA}$, as this is regarded the length scale of atomic interactions.

\mysubsubsection{Data noise}
A configuration's total quantum mechanical energy and its derivatives can be extrapolated to exact numerical values, provided the various convergence parameters of the applied quantum mechanical method are used appropriately. However, we should still regard these as noisy observations when trying to fit a model, for the following reasons:
\begin{enumerate}[(i)]
\item the separation into a sum of local contributions is an approximation,
\item our model is additive over various contributions with unknown individual ratios,
\item our model employs a finite cutoff,
\item the quantum mechanical calculations may not be fully converged.\footnote{This could be an advantage, as we do not \emph{need} fully converged quantum mechanical data.}
\end{enumerate}
Thus we  modify our model in \refeq{eq:totale_full_mbe} to include a Gaussian noise $\nu_E$,
$P(\nu_E) = \mathrm{Normal}(\nu_E; 0, \sigma_E)$
\begin{equation}\label{eq:totale_full_mbe_noise}
E=\sum_\alpha \sum_i \varepsilon^\alpha_i + \nu_E
\end{equation}
and similarly, derivative quantities are modelled as
\begin{equation}\label{eq:de_full_mbe_noise}
\frac{\partial E}{\partial \xi_k}=\frac{\partial\sum_\alpha \sum_i \varepsilon^\alpha_i}{\partial \xi_k} + \nu_\xi
\end{equation}
where $P(\nu_\xi) = \mathrm{Normal}(\nu_\xi; 0, \sigma_\xi)$. As a consequence, \refeq{eq:TE_covariance} is  modified to give the covariance of \emph{observed} total energies
\begin{equation}
\langle E_N E_M \rangle = \sigma_w^2 \sum_{i \in N} \sum_{j \in M} C(\mathbf{d}_i,\mathbf{d}_j) + \sigma_E^2 \delta_{NM}
\textrm{,}
\end{equation}
and the covariance of observed derivatives becomes
\begin{equation}
\frac{\partial^2 \langle E_N E_M \rangle}{\partial \xi_k \partial \chi_l} = 
\sigma_w^2 \sum_{i \in N} \sum_{j \in M} \frac{\partial \mathbf{d}_i^\top}{\partial \xi_k}  (\nabla_{\mathbf{d}_i} C(\mathbf{d}_i,\mathbf{d}_j) \nabla_{\mathbf{d}_j}^\top)
\frac{\partial \mathbf{d}_j}{\partial \chi_l}+ \sigma_\xi^2 \delta_{NM} \delta_{\xi_k \chi_l}
\end{equation}

\mysubsubsection{Sparsification}
It is easy to see that computing covariance matrices and vectors can become quite expensive, especially if derivative quantities are also included. This, combined with the assumption that atomic neighbourhood environments are often repetitious, leads to the idea that sparse Gaussian processes might be applied. Sparsity is a central concept in machine learning, and sparse Gaussian processes are described in detail by, for example, Qui\~nonero-Candela and Rasmussen,\cite{QuinoneroCandela:2005wp} or Snelson and Ghahramani\cite{Snelson:2006vi}. In our adaptation of sparsification, we use  {\em\ representative} atomic neighbourhood environments, or pairs and triplets etc. The model is built using all observations in the dataset and it can be regarded as a projection onto a subset of data points, the sparse representation.

Let us consider a set of configurations, each of which contains an arbitrary number of atoms and the corresponding set of total energies, derivatives or both. The observables are collected in the vector $\mathbf{t}$.
We select a set of environments, the sparse set $S$, and compute the covariance matrices:
\begin{equation}
(\mathbf{C}_{SS})_{ss'} = \langle \varepsilon_s \varepsilon_{s'} \rangle \textrm{, where } s,s' \in S
\textrm{,}
\end{equation}
\begin{equation}
(\mathbf{C}_{ST})_{st} = \langle \varepsilon_s E_t \rangle \textrm{,}
\end{equation}
where $s \in S$ and $t$ is an  index of  total energies in $\mathbf{t}$, and
\begin{equation}
(\mathbf{C}_{ST})_{s\tau} = \langle \varepsilon_s \frac{\partial E_t}{\partial \xi_k} \rangle \textrm{,}
\end{equation}
where $s \in S$ and $\tau$ denotes derivative observables in $\mathbf{t}$. The predicted value at an arbitrary atomic neighbourhood environment $\mathbf{d}_*$ can be calculated from
\begin{equation}
\varepsilon_* (\mathbf{d}_*)= \mathbf{k}_*^\top (\mathbf{C}_{SS}+\mathbf{C}_{ST} \Lambda_{TT}^{-1} \mathbf{C}_{TS})^{-1}
\mathbf{C}_{ST} \Lambda_{TT}^{-1} \mathbf{t}
\textrm{,}
\end{equation}
where $(\mathbf{k}_*)_s=  \langle \varepsilon_* \varepsilon_{s} \rangle$, and $\Lambda_{TT}$ is a diagonal matrix, where each diagonal element is $\sigma^2_E$ or $\sigma^2_\xi$, depending on the type of observable. As configurations may contain different numbers of atoms, we scale $\sigma^2_E$ accordingly. Note that the part multiplying $\mathbf{k}_*$ from the right is precomputed at the training stage, so only $\mathbf{k}_*$ needs to be computed for each prediction, and this scales linearly with the number of sparse points (and not with the total number of original data points!). Derivatives of $\varepsilon_*$ are readily available analytically, using the appropriate covariance functions.
In practice, we found that the sparse covariance matrix $\mathbf{C}_{SS}$ should be regularised by adding a small positive constant $\sigma_\textrm{jitter}$ to the diagonal values. The numerical value of the constant should be as small as possible, without compromising the positive definiteness of  $\mathbf{C}_{SS}$. Normally $\sigma_\textrm{jitter}$ is 6-9 orders of magnitude less than the diagonal elements.

\begin{figure}
\begin{center}
\includegraphics[width=0.9\columnwidth,keepaspectratio=true]{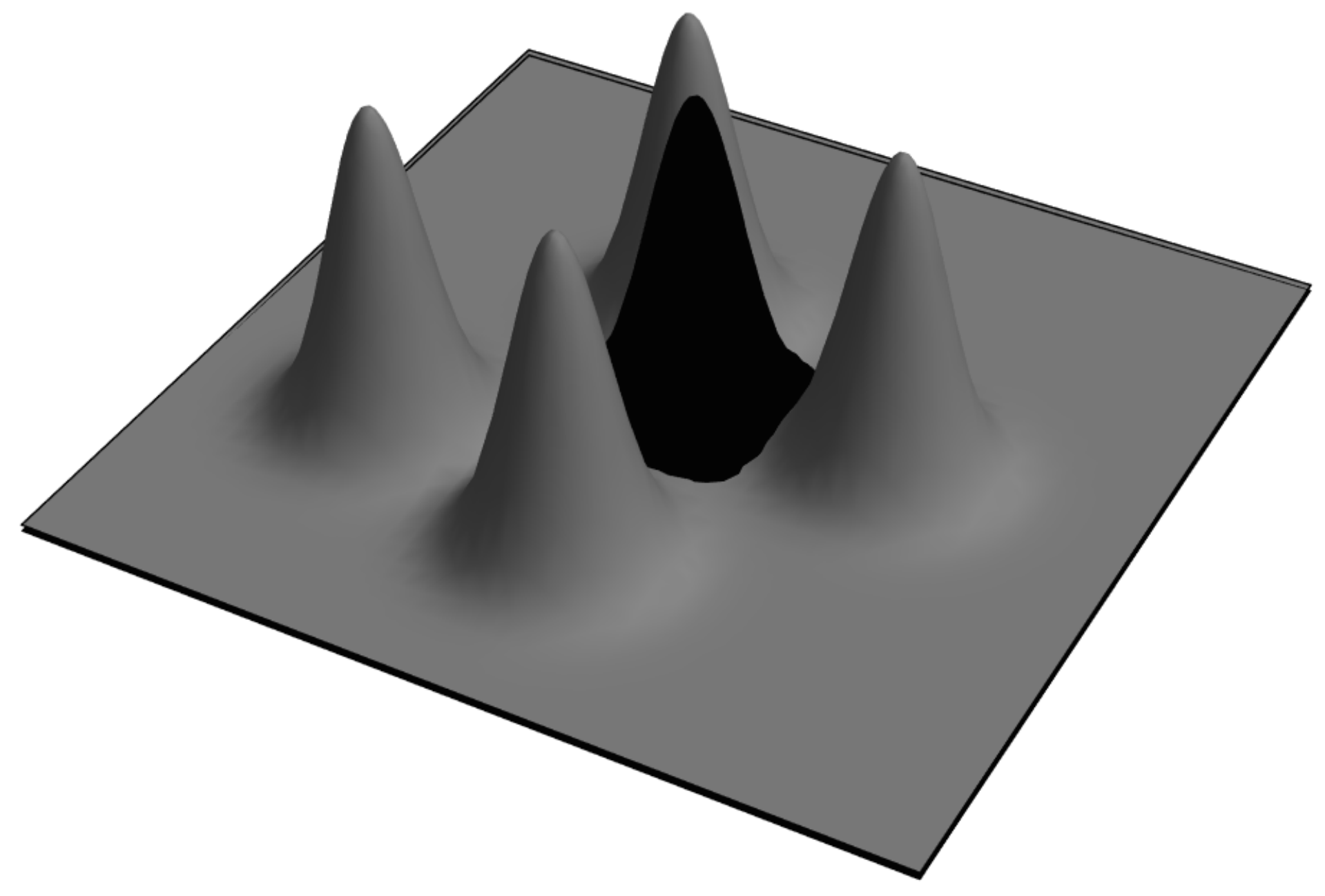}
\end{center}
\caption{\label{fig:SOAP_density} A two-dimensional illustration of the atomic neighbour density function used in SOAP. The white circle represents the radial cutoff distance. }
\end{figure}

\mysubsection{Descriptors}
The success of applying machine learning techniques to fit potential energy surfaces depends to a large extent   on  representing the atomic environments appropriately. Transformations of atomic positions to which the local energy  is invariant, \ie \ rotation and inversion of an environment about its centre, and permutation of identical atoms should be explicitly built in. We presented a detailed study on representing chemical environments elsewhere\cite{Bartok:2013cs} in which we focussed on atom-centred neighbourhood environments. Here we describe a few other types of descriptors.
\mysubsubsection{Pairs and triplets}
Pairs of atoms are simply described by the  distance between them, but in case of triplets the distances need to be symmetrised. If atoms $j$ and $k$ form a triplet with $i$ as the central atom, a possible descriptor can be the vector
\begin{equation}
\label{eq:triplet_descriptor}
[r_{ik}+r_{ij},(r_{ik}-r_{ij})^2, r_{jk}] \textrm{.}
\end{equation}
As we mentioned earlier, the covariance function must be augmented by a cutoff function. We use $f_\textrm{cut}(r_{ij})$ for the pair terms, and in case of triplets we use $f_\textrm{cut}(r_{ij}) f_\textrm{cut}(r_{ik})$.
\mysubsubsection{Water dimers}
It is clear that our approach to symmetrise distances in case of three-body descriptors will be overly complicated if we attempt to apply it on more than a couple of atoms. For example, the potential energy surface of water can be modelled very accurately using a many-body expansion of interactions between water molecules.\cite{Gillan:2013hq,Medders:2014bp} The two-body term in the expansion necessitates a descriptor for the water-water dimer, for which we used the pairwise distances between the constituent atoms, 15 in total. However, this descriptor in this  form is not invariant to permuting atoms of the same element. If exchange of hydrogen atoms between different molecules is not permitted, the following permutations $\hat{P}$ that operate on the order of atoms must be taken in account: 
\begin{enumerate}[(i)]
\item swaps of water molecules in the dimer (2)
\item exchange of hydrogen atoms within each molecule ($2 \times 2$),
\end{enumerate}
so 8 in total. Instead of modifying the descriptor, we enforced permutation symmetry at the level of the kernel function. If we take an arbitrary kernel function, $C(\mathbf{d},\mathbf{d}')$, that takes vector arguments, we can generate a permutational invariant kernel as
\begin{equation}
C'(\mathbf{d},\mathbf{d}') = \sum_{\hat{P}} C(\mathbf{d},\hat{P}\mathbf{d}')
\textrm{,}
\end{equation}
which must be normalised\cite{Williams:2007vz}:
\begin{equation}
C''(\mathbf{d},\mathbf{d}') = \frac{C'(\mathbf{d},\mathbf{d}')}{\sqrt{C'(\mathbf{d},\mathbf{d})}\sqrt{C'(\mathbf{d}',\mathbf{d}')}}
\textrm{.}
\end{equation}
We used the squared exponential as our starting kernel in the case of water molecules.
\mysubsubsection{SOAP}
We note that our previously introduced \cite{Bartok:2013cs} kernel based on ``Smooth Overlap of Atomic Positions'' may be interpreted from the function-space view we used throughout  this manuscript. We represent the atomic neighbourhood of atom $i$ by the neighbourhood density function (for illustration, see \reffig{fig:SOAP_density})
\begin{equation}
\label{eq:SOAP_density}
\rho_i(\mathbf{r}) \equiv \sum_j^\textrm{neigh.} \exp \left(- \frac{| \mathbf{r} -\mathbf{r}_{ij} |^2}{2 \sigma_\textrm{atom}^2} \right)\textrm{,}
\end{equation}
and $\varepsilon_i$, the atomic energy of atom $i$ can then be regarded as a functional of $\rho_i$
\begin{equation}
\varepsilon_i = \varepsilon[\rho_i] = \int \! w(\mathbf{r}) \rho_i(\mathbf{r}) \dd \mathbf{r}
\end{equation}
where the prior distribution of the weights is Gaussian, so
\begin{equation}
\langle w(\mathbf{r}) w(\mathbf{r}') \rangle = \delta(\mathbf{r}-\mathbf{r}') \sigma^2_w
\textrm{,}
\end{equation}
resulting in the covariance of two atomic energies
\begin{equation}
C(\rho_i,\rho_j) =
\langle \varepsilon_i \varepsilon_j \rangle = \left \langle 
\int \! w(\mathbf{r}) \rho_i(\mathbf{r}) 
w(\mathbf{r'}) \rho_j(\mathbf{r'}) \,\dd \mathbf{r} \, \dd \mathbf{r'} \right\rangle =
\sigma^2_w \int  \rho_i(\mathbf{r})  \rho_j(\mathbf{r}) \,\dd \mathbf{r}
\textrm{.}
\end{equation}
It is useful to note here that if $C$ is a valid kernel, then $|C|^p$ is also valid\cite{Williams:2007vz}. This covariance function $|C|^p$  is not invariant to rotations, but we may convert it in a similar fashion to what we did in the case of the water dimer:
\begin{equation}
C'(\rho_i,\rho_j) = \int \! |C(\rho_i,\hat{R}\rho_j)|^p\dd \hat{R}
\textrm{,}
\end{equation}
which must then be normalised, so the final result is
\begin{equation}
\label{eq:normalise_soap}
C''(\rho_i,\rho_j) = \frac{C'(\rho_i,\rho_j)}{\sqrt{C'(\rho_i,\rho_i)}\sqrt{C'(\rho_j,\rho_j)}}
\textrm{.}
\end{equation}

In practice, we evaluate the SOAP kernel numerically by first expanding \refeq{eq:SOAP_density} in a basis set
\begin{equation}
\label{eq:soap_radialexp}
\rho_i(\mathbf{r}) = \sum_{nlm} c^{(i)}_{nlm} g_n(r) Y_{lm}(\hat{\mathbf{r}})
\textrm{,}
\end{equation}
where $c^{(i)}_{nlm}$ are the expansion coefficients corresponding to atom $i$, $\{g_n(r)\}$ is  an arbitrary set of orthonormal radial basis functions, and $Y_{lm}(\hat{\mathbf{r}})$ are the spherical harmonics. We form descriptors from the coefficients by computing the power spectrum elements
\begin{equation}
p^{(i)}_{nn'l} \equiv \frac{1}{\sqrt{2 l + 1}}\sum_m c^{(i)}_{nlm} (c^{(i)}_{n'lm})^*
\textrm{,}
\end{equation}
and the rotationally invariant covariance of atoms $i$ and $j$ is given by 
\begin{equation}
\label{eq:kernel_power}
C'(\rho_i,\rho_j) = \sum_{n,n',l} p^{(i)}_{nn'l} p^{(j)}_{nn'l}
\textrm{,}
\end{equation}
which we normalise according to \refeq{eq:normalise_soap}. The normalisation step is equivalent to normalising the vector elements $p^{(i)}_{nn'l}$, so $C''$ is, in fact, a dot-product kernel of vectors $\mathbf{p}^{(i)}/|\mathbf{p}^{(i)}|$ and $\mathbf{p}^{(j)}/|\mathbf{p}^{(j)}|$. Note that it is often useful to raise $C''$ to a power $\zeta > 1$, in order to sharpen the difference between atomic environments. To see the details of the above results, we refer the reader to our earlier work.\cite{Bartok:2013cs,Szlachta:2014jh}.

\mysection{SOFTWARE}
\sloppy
The ideas presented in the Methodology section are implemented in the QUIP package, which can be downloaded from the git repository at \verb+https://github.com/libAtoms/QUIP+. Code related to GAP prediction can be obtained under an non-commercial licence from 
\verb+http://www.libatoms.org/gap/gap_download.html+. Users who wish to use the training code should contact the corresponding author.

QUIP is a  molecular simulation sandbox written in object-oriented FORTRAN95/2003, with interfaces to python (compatible with ASE), and various other simulation packages, such as LAMMPS, CP2K,  CASTEP, and others. QUIP is essentially a collection of objects and interfaces that contain and manipulate atomic configurations and interatomic potentials.

The GAP implementation provides over 20 different descriptors, which can be used with two types of covariance functions, the squared exponential 
\begin{equation}
C(\mathbf{d}_i,\mathbf{d}_j)=\delta^2 \exp \left( -\frac{1}{2}\sum_\alpha \left( \frac{d_{i\alpha}-d_{j\alpha}}{\theta_\alpha}\right)^2 \right)
\end{equation}
and the polynomial kernel
\begin{equation}
C(\mathbf{d}_i,\mathbf{d}_j) = \delta^2 (\mathbf{d}_i \cdot \mathbf{d}_j + \sigma_0^2)^\zeta
\textrm{.}
\end{equation}

\fussy

To demonstrate the training, we provide a simple example, where we train GAP to reproduce the well-known Stillinger-Weber (SW) potential\cite{STILLINGER:1985vx} for silicon. We used a database of $\textrm{Si}_n$ silicon clusters, $n=7\ldots13$, sampled from a 2000 K molecular dynamics simulation. We used 600 configurations in total, with total energies and forces calculated using the SW potential. We used a combination of two- and three-body interactions, with a cutoff of 4.1~\AA.
We used the command line \\
\verb+ teach_sparse at_file+\verb+=+\verb+data_Si_SW.xyz descriptor_str={ \ + \\
\verb+ distance_2b cutoff=4.1 n_sparseX=250 covariance_type=ard_se theta_fac=0.5: \ + \\
\verb+ angle_3b cutoff=4.1 n_sparseX=500 covariance_type=ard_se theta_fac=0.5} \ + \\
\verb+ default_sigma={0.001 0.05 0.01} sparse_jitter=1.0e-8 e0=0.0 + \\
Table \ref{tbl:teach_sparse} summarises the command line arguments used in this example. 
\begin{table}[h]
\begin{tabular}{|l|l|}\hline
\verb+at_file+ & contains the database configurations, in concatenated XYZ files\\ \hline
\verb+descriptor_str+ & parameters of descriptor(s)\\ \hline
\verb+default_sigma+ & the assumed standard deviation of the errors, $\{\sigma_\textrm{energy}\;\sigma_\textrm{force}\;\sigma_\textrm{viral}\}$\\ \hline
\verb+sparse_jitter+ & regularisation of the sparse covariance matrix, $\sigma_\textrm{jitter}$\\ \hline
\verb+e0+ & baseline of atomic energies\\ \hline
\end{tabular}
\caption{\label{tbl:teach_sparse}Overview of the basic command line arguments of \texttt{teach\_sparse}.}
\end{table}
The command line argument \verb+descriptor_str+ contains the parameters of the descriptor, which depend on the type. We  define the number of sparse points \verb+n_sparseX+ and the type of covariance function. \verb+theta_fac+ is the simplest way to control the $\theta_\alpha$ length-scale parameters in the squared exponential covariance function: the range of the descriptor values in each dimension $\alpha$ is scaled by the constant \verb+theta_fac+. More descriptors can be concatenated, separated by the \verb+:+ symbol, resulting in the  fitting of a model that is the sum terms each based on one descriptor.

It is possible to specify an existing QUIP potential as a baseline, so energies/forces/virials are subtracted from the target values before fitting. These are added back automatically when the potential is called. Naturally, the baseline can be another GAP, resulting in hierarchical models with arbitrary level of recursion.

\begin{figure}
\begin{center}
\includegraphics[width=0.9\columnwidth,keepaspectratio=true]{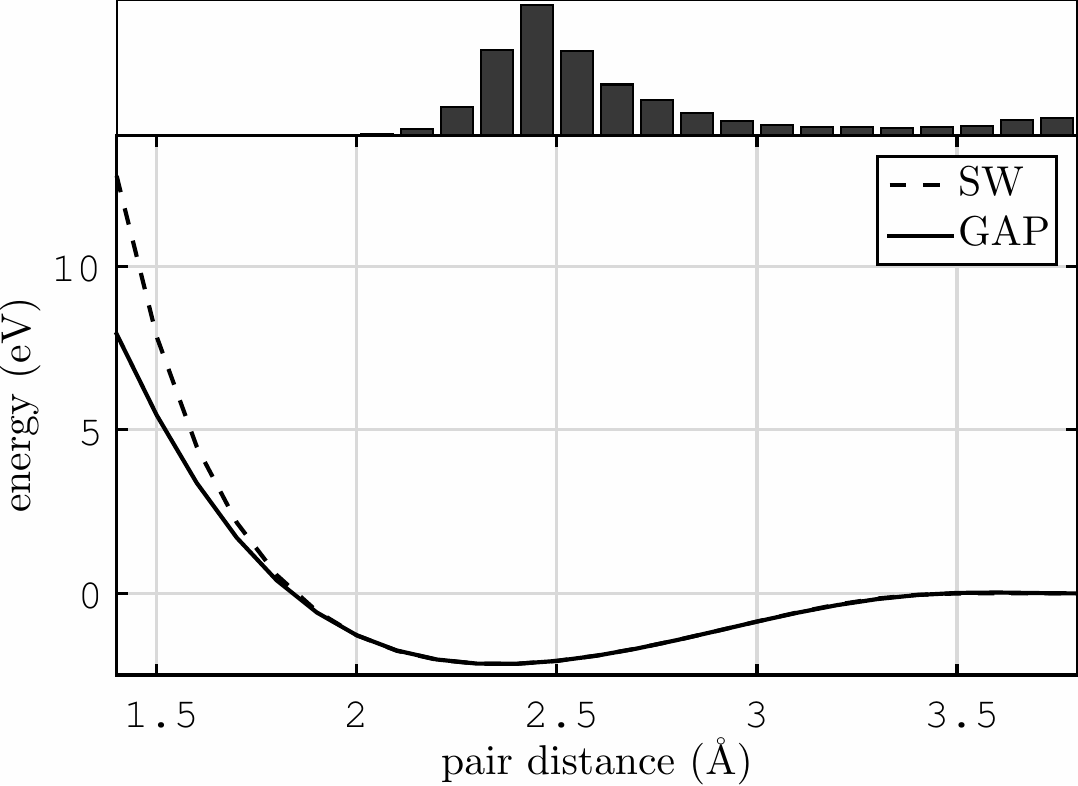}
\end{center}
\caption{\label{fig:SW_2body} The two-body term in the Stillinger-Weber potential for silicon. The dashed line represents the analytical answer, the solid line shows the GAP fit. The histogram above corresponds to the input data to the fit.}
\end{figure}

\begin{figure}
\begin{center}
\includegraphics[width=0.9\columnwidth,keepaspectratio=true]{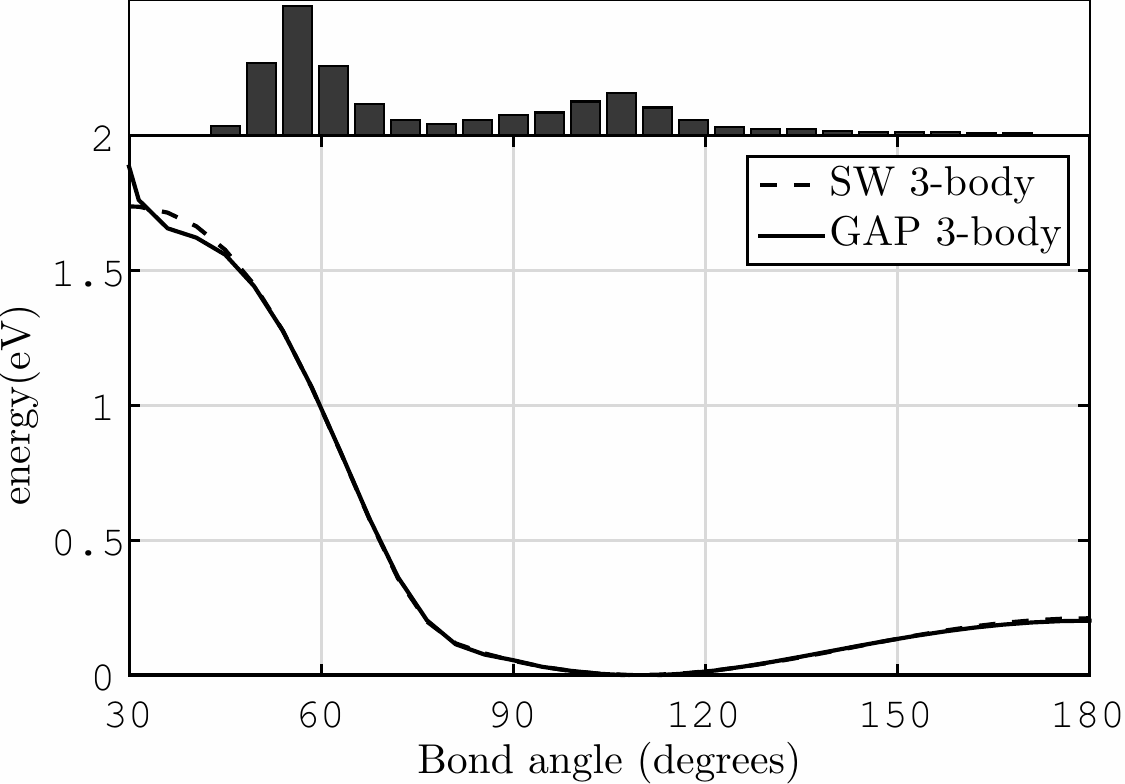}
\end{center}
\caption{\label{fig:SW_3body} The three-body term in the Stillinger-Weber potential for silicon. We fixed the two neighbours at 2.5~\AA\ and 2.8~\AA\ and varied the bond angle and  plotted the sum of all three interactions between the three atoms. The dashed line represents the analytical answer, the solid line shows the GAP fit. The histogram above represents the input data to the fit. }
\end{figure}

In the above example, it is possible to check whether the GAP model was able to recover both terms of the target model, as they are available analytically.  We emphasise that the fitting uses total energies and forces only, so the machine learning algorithm has  to infer the separate two- and three-body terms from this convoluted data. To show the quality of the fit, we plot the pair potential in \reffig{fig:SW_2body} and the angle term in \reffig{fig:SW_3body} for both the original SW and the fitted model. The agreement is rather good in both cases, except at the edges of the range where there was no input data.

The potential file generated by \verb+teach_sparse+ can be used to compute the total energies and similar quantities  of arbitrary configurations. For example \\
\verb+ eval at_file+\verb+=+\verb+data_Si_SW.xyz param_file+\verb+=+\verb+gp.xml init_args={IP GAP} e f+\\
will compute the total energies (``\verb+e+'') of the configurations stored in \verb+data_Si_SW.xyz+ as well as the atomic forces (``\verb+f+'') using the fitted GAP model from the file \verb+gp.xml+. Another use of \verb+eval+ is to compute and print the descriptor vectors for any descriptor type implemented in QUIP. For example, \\
\verb+ eval at_file+\verb+=+\verb+data_Si_SW.xyz descriptor_str={angle_3b cutoff=4.1}+\\
prints all three-body descriptors, in this case the descriptors defined in \refeq{eq:triplet_descriptor}.

%
%


\mysubsection{ACKNOWLEDGMENTS}
A.P.B. is supported by a Leverhulme Early Career Fellowship and the Isaac Newton Trust. We would like to thank our referees for their comments during the revision process.

\clearpage


\bibliography{gap-tutorial}   


\clearpage

%

\end{document}